# CDEX Dark Matter Experiment: Status and Prospects

**Hao Ma[1], Yang Chen[1], Qian Yue[1], Li Wang[1,2], Tao Xue[1], Zhi Zeng[1], Kejun Kang[1], Jianping Cheng[1,2], Yuanjing Li[1], Jianmin Li[1], Yulan Li[1]**

1. Key Laboratory of Particle and Radiation Imaging (Ministry of Education) and Department of Engineering Physics, Tsinghua University, Beijing 100084, China
2. Beijing Normal University, Beijing 100875, China

E-mail: mahao@tsinghua.edu.cn, yueq@tsinghua.edu.cn.

**Abstract**. The China Dark Matter Experiment (CDEX) aims at direct searches of light Weakly Interacting Massive Particles (WIMPs) at the China Jinping Underground Laboratory (CJPL) with an overburden of about 2400m rock. Results from a prototype CDEX-1 994 g p-type Point Contact Germanium(pPCGe) detector are reported. Research programs are pursued to further reduce the physics threshold by improving hardware and data analysis. The CDEX-10 experiment with a pPCGe array of 10 kg target mass range is being tested. The evolution of CDEX program into "CDEX-1T Experiment" with ton-scale germanium detector arrays will also be introduced in this study.

## 1. Introduction to CDEX

Various astronomical and cosmological evidences indicate the existence of dark matter, which contributes about one quarter of the energy density of the Universe. As the most desirable candidates of the dark matter, weakly interacting massive particles(WIMPs) could interact with nuclei of normal matter via elastic scattering, the recoil energy of which can be detected in extreme low background experiments at deep underground laboratories[1]. China Jinping Underground Laboratory(CJPL), located in Sichuan Province southwest of China, is an ideal site to implement dark matter searches with a rock overburden of 2400 meters. Over the past decade, p-type point contact germanium detectors(pPCGe), sensitive to sub-keV recoil energy, were used to search light WIMPs with mass 1GeV to 10 GeV[2,3]. The China Dark Matter Experiment (CDEX) aims at direct detection of light WIMPs using pPCGe and started to run at CJPL in 2010[4]. In this study, results from first generation of CDEX with two ~1kg pPCGe detectors and current phase with an array of 10 kg pPCGe detectors will be reported. The future plan of CDEX-1T with ton-scale pPCGe arrays will also be presented.

## 2. Status and prospects of CDEX

The first generation of CDEX (CDEX-1) adopted single-element ~1kg mass pPCGe detector with a low background well-type NaI(Tl) anti-Compton detector. The detectors were surrounded by combined shielding structure made of oxygen-free high-conductivity (OFHC) copper, borated polyethylene (PE(B)), and lead, as depicted in Fig 1. In addition, the detectors and OFHC copper bricks were placed in a plastic bag with a nitrogen gas flush to exhaust radon. The entire setup was installed in a room, with one-meter-thick polyethylene floor, roof and walls, at CJPL[5].

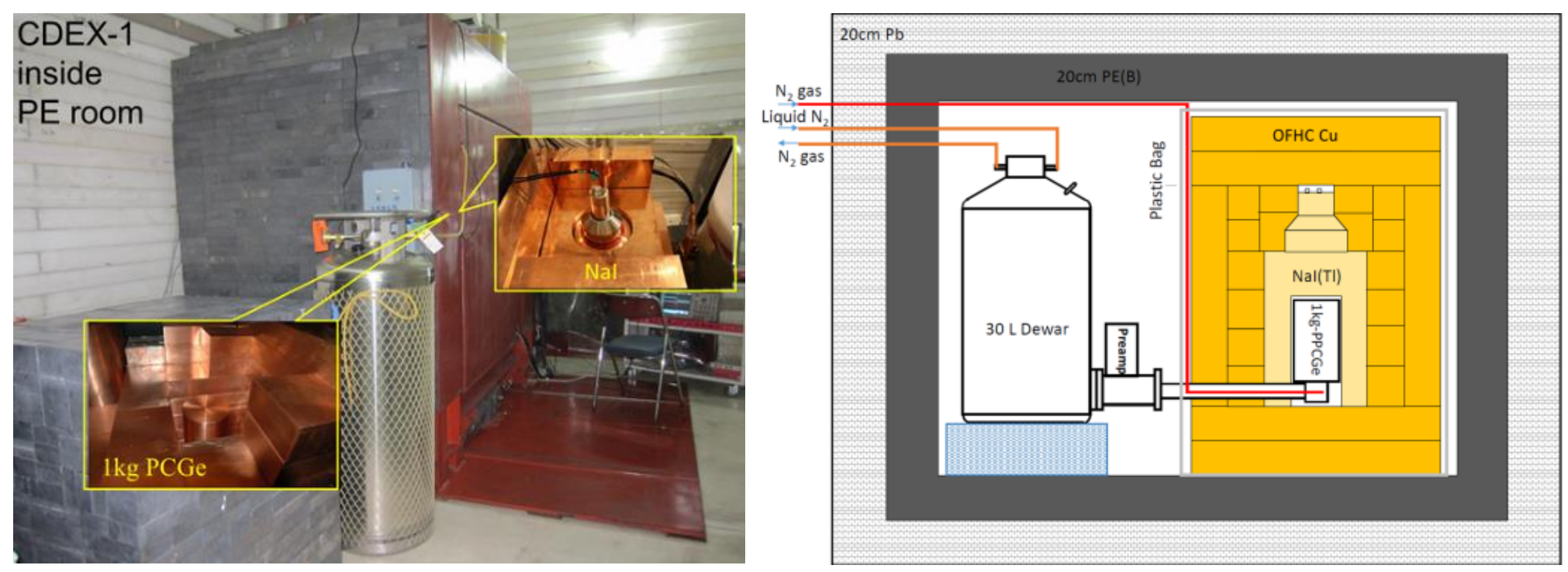

**Fig 1.** The photo and schematic view of CDEX-1 experiment. OFHC copper, borated polyethylene and lead were used as a passive shielding, and NaI(Tl) as a veto detector. Nitrogen gas evaporating from dewar was flushed into a plastic bag surrounding the copper bricks to exhaust radon.

Based on 53.9 kg-days of data[6], first result from CDEX-1 was reported with an energy threshold of 475 eVee, and WIMP-nucleus coherent elastic scattering at WIMP mass of 6–20 GeV/$c^2$ were probed and excluded, shown in Fig2. The dark matter region favored by CoGeNT experiment[3], also using pPCGe detector technique, was excluded. Subsequent data taking was implemented[7], a longer exposure of 335.6 kg·day further improved sensitivity ~2 times with a flat background level reduced to 3 cpkkd (counts/keV/kg/day), shown in Fig3. A more stringent limit to spin-independent and spin-dependent WIMP-nucleon scattering cross section was set for a range of light WIMP mass below 10 GeV/$c^2$ shown in Fig2. The results represent the most sensitive measurements made with the pPCGe detector.

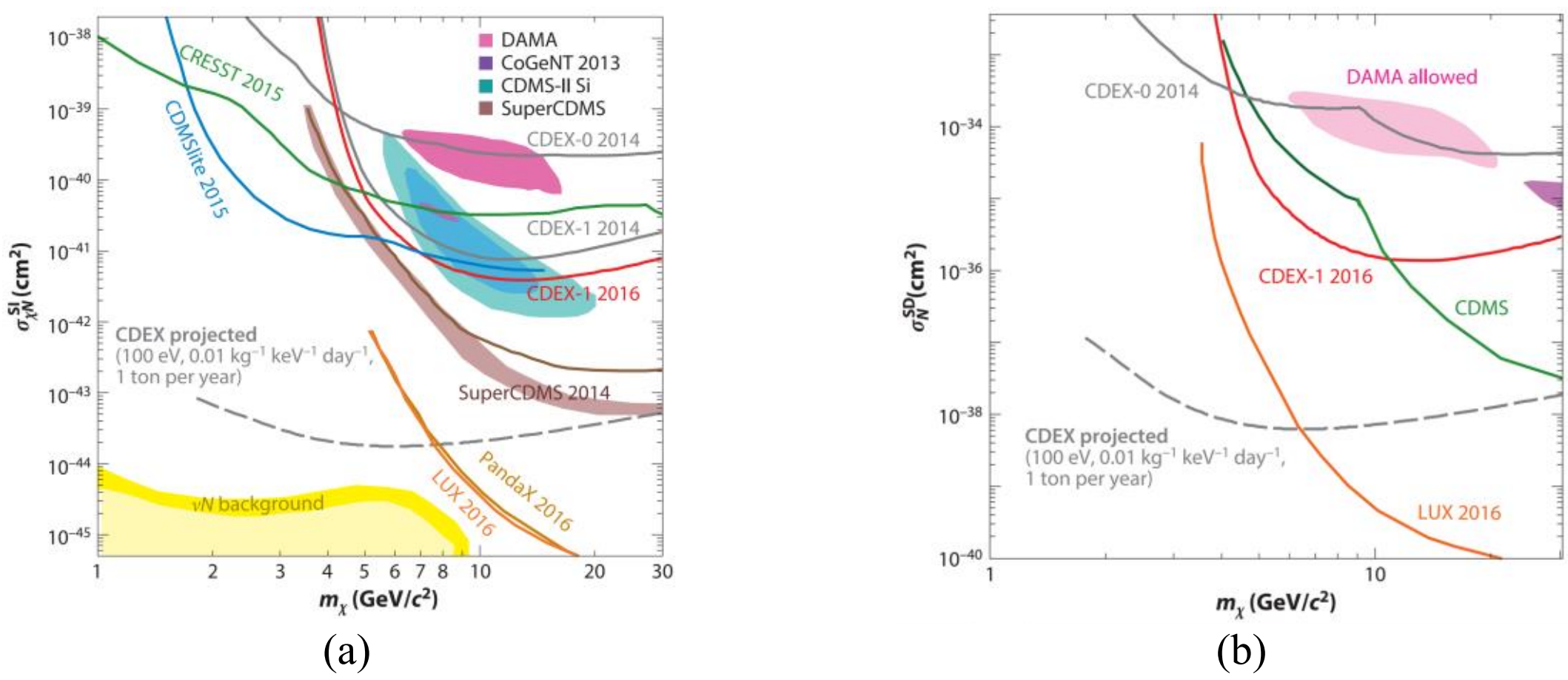

(a) (b)

**Fig 2.** Regions in couplings versus the WIMP mass ($m_\chi$) parameter space probed and excluded by the CDEX-1 experiment with 335.6 kg·day of exposure, along with comparisons with other benchmark results[8–10]. (a) Spin-independent WIMP–nucleon couplings. (b) Spin-dependent WIMP–neutron couplings[7].

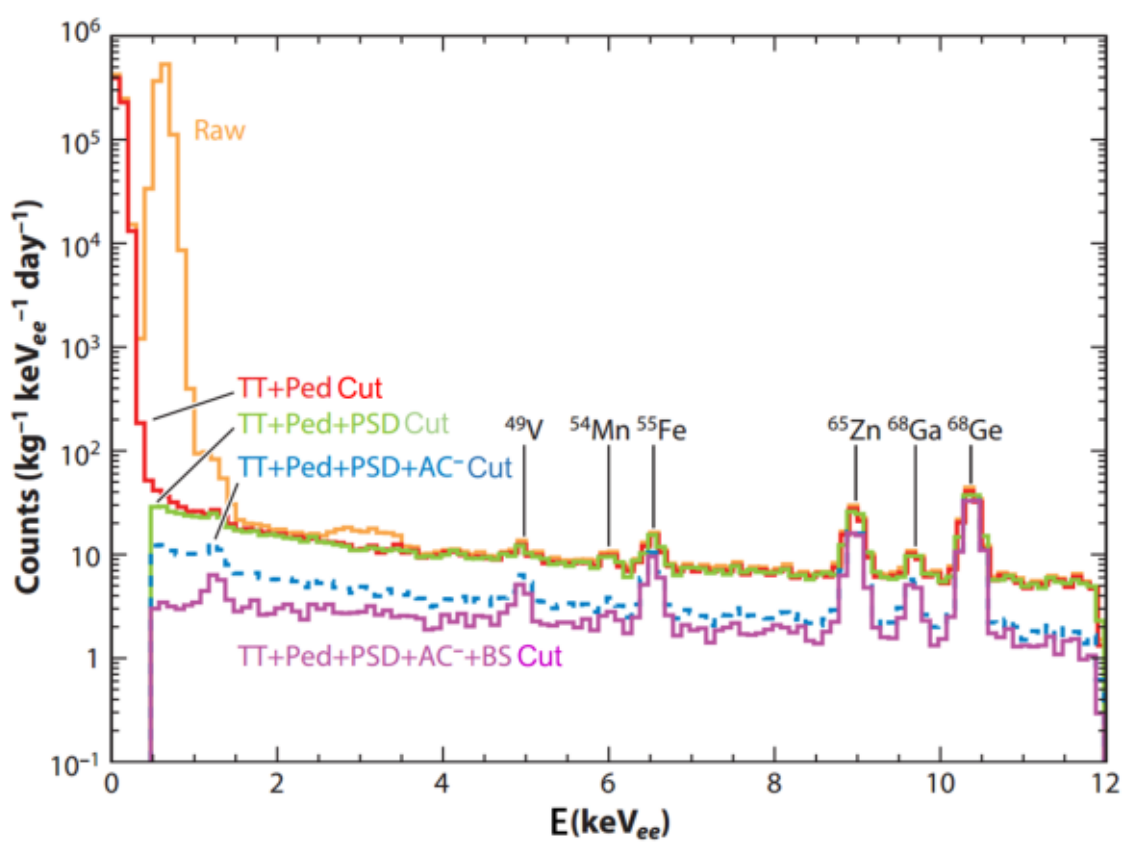

**Fig 3.** Background spectra of CDEX-1 experiment with an exposure of 335.6 kg·day (modified from [7]). Spectra after various cuts of candidate events: basic cuts (TT+Ped+PSD), anti-Compton cut (AC⁻), and bulk/surface cut (BS). The peaks correspond to internal X-ray emission from cosmogenic isotopes with relatively long half-lives.

The result on $^{76}$Ge neutrino-less double beta decay(0υββ) from CDEX-1 natural germanium crystal was also analyzed with a exposure of 304 kg·d[11]. The average event rate obtained was about 0.012 cpkkd over the interested 2.039 MeV energy range. The half-life of $^{76}$Ge 0υββ derived based on this result is $T^{0\nu}_{1/2}>6.4\times10^{22}$yr(90%C.L.), with an upper limit on the effective Majorana-neutrino mass of 5.0 eV, shown in Fig 4.

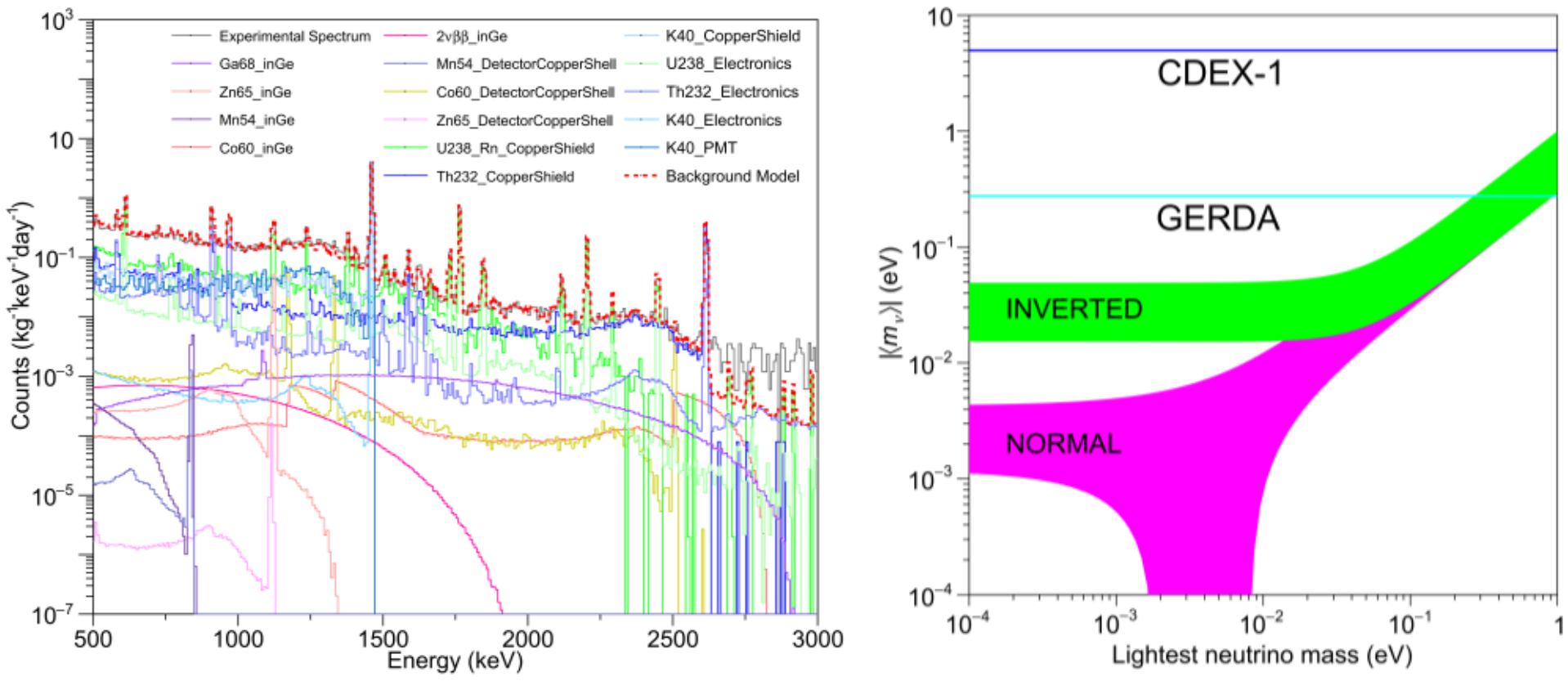

**Fig 4.** Background model of CDEX-1(500keV to 3MeV) and upper limits on the effective Majorana-neutrino mass compared with that from GERDA Collaboration[11].

For dark matter searches, CDEX experiment tries to lower both energy threshold and background as much as possible. Based on the results from first pPCGe detector(CDEX-1A), the point contact electrode of a new detector(CDEX-1B) was further improved with a lower noise JFET and updated pre-amplifier. The updated detector achieved the energy threshold of ~180eVee. The year-long data set of CDEX-1B is currently being analyzed, and the new physical results are expected to be reported soon.

The long-term physical goal of CDEX project is a ton-scale germanium experiment (CDEX-1T) searching for dark matter and 0νββ. The experimental setup will adopt germanium detector arrays deploying in a liquid nitrogen tank with 13 meters in diameter and 13 meters in height, shown in Fig5. The liquid nitrogen acts as cooling medium for germanium arrays and passive shielding material against ambient radioactivity simultaneously. The tank will be located in a pit with a diameter of 18 meters in and a height of 18 meters in the Hall C of the expansion project of CJPL (called CJPL-II, and existing part of CJPL called CJPL-I)[4], as depicted in Fig5.

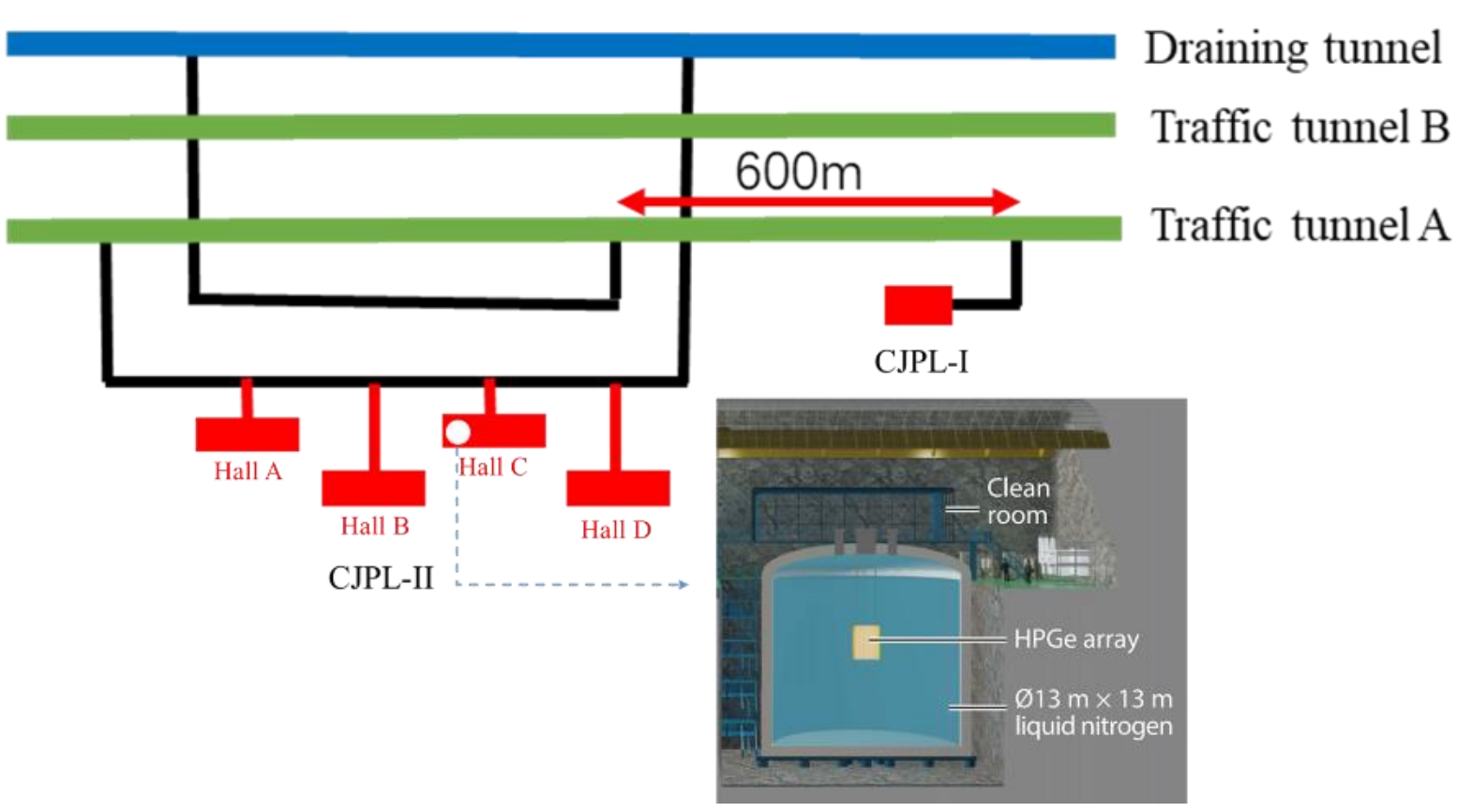

**Fig 5.** Schematic layout of CJPL and Conceptual design of CDEX-1T(modified from [4]).

Toward CDEX-1T plan, the so-called CDEX-10 prototype system has been established with germanium detectors, total mass of ~10kg, within an array structure deploying in a much smaller liquid nitrogen tank (1.5m in diameter and 1.9m in height). OFHC copper and lead bricks are still used as exterior shielding materials. The CDEX-10 setup is placed adjacent to CDEX-1 in the polyethylene room at CJPL-I. CDEX-10 acts as a platform to study the many issues toward CDEX-1T, involving scaling up in detector mass, lowering energy threshold and reducing background. The projected sensitivity is superimposed in Fig3. A 60kg·d data set is being analyzed to study the characteristics, e.g. energy resolution, threshold and background, of array detectors. In addition, we also study cosmogenic isotopes inside the germanium crystal and copper by Monte Carlo simulation and experiments simultaneously.

## 3. Summary

The status and prospects of CDEX dark matter experiment were presented. CDEX-1A achieved the energy threshold of 475 eVee and disfavored the dark matter region reported by CoGeNT experiment. Improved with lower detector capacitance, CDEX-1B got an energy threshold of ~180eVee and is under analysis with a year-long data set. Toward the CDEX-1T roadmap, CDEX-10 was carried out at CJPL-I as a prototype to study characteristics of array detectors, performance of liquid nitrogen on cooling and shielding, and background model. The liquid nitrogen cryostat of CDEX-1T with a testing platform will be established at CJPL-II at the end of next year. A new customized and updated pPCGe array detectors will arrive at CJPL and be tested to evaluate the performance of the entire system.

## Acknowledgement

This work is supported by the National Key Research and Development Program of China (No. 2017YFA0402200 and 2017YFA0402201), the National Natural Science Foundation of China (Nos.11175099, 11275107, 11475117, 11475099, 11475092 and 11675088), the National Basic Research Program of China (973 Program) (2010CB833006), and the grants from the Tsinghua University Initiative Scientific Research Program (No.20121088494, 20151080354).